\documentclass{article}

\usepackage{PRIMEarxiv}

\usepackage[utf8]{inputenc}
\usepackage[T1]{fontenc}   
\usepackage{url}          
\usepackage{booktabs}     
\usepackage{amsfonts}     
\usepackage{nicefrac}     
\usepackage{microtype}     
\usepackage{lipsum}
\usepackage{graphicx}
\usepackage{amssymb}
\usepackage{booktabs,chemformula}
\usepackage{dutchcal}
\usepackage[defaultmathsizes]{mathastext}

\usepackage{wasysym}
\usepackage{longtable}
\usepackage{amsmath}
\usepackage{booktabs}
\usepackage{multicol}
\usepackage{xtab}
\usepackage{caption}
\usepackage{natbib}
\usepackage{url}

\usepackage{breakurl}

\usepackage[breaklinks,hidelinks]{hyperref}
\usepackage{pdflscape}   
  
\title{Rapid Shear Capacity Prediction of TRM-Strengthened Unreinforced Masonry Walls through Interpretable Machine Learning using a Web App
 
}
\author{
  Petros C. Lazaridis, Athanasia K. Thomoglou\\
  Department of Civil Engineering\\
  Democritus University of Thrace \\
  Xanthi, Greece\\
  \texttt{\{plazarid, athomogl\}@civil.duth.gr} \\
}

\begin{document}
\setcitestyle{numbers}
\maketitle

\begin{abstract}
The presented study aims to provide an efficient and reliable tool for rapid estimation of the shear capacity of a TRM-strengthened masonry wall. For this purpose, a data-driven methodology based on a machine learning system is proposed using a dataset constituted of experimental results selected from the bibliography. The outlier points were detected using Cook's distance methodology and removed from the raw dataset, which consisted of 113 examples and 11 input variables. In the processed dataset, 17 Machine Learning methods were trained, optimized through hyperparameter tuning, and compared on the test set. The most effective models are combined into a voting model to leverage the predictive capacity of more than a single regressor. The final blended model shows remarkable predicting capacity with the determination factor ($R^2$)equal to 0.95 and the mean absolute percentage error equal to 8.03\%. In sequence, machine learning interpretation methods are applied to find how the predictors influence the target output. $A_m$, $f_t$, and $n\cdot t_f$ were identified as the most significant predictors with a mainly positive influence on the shear capacity. Finally, the built ML system is employed in a user-friendly web app for easy access and usage by professionals and researchers. 
\end{abstract}
\keywords{Shear Capacity \and TRM \and Machine Learning \and Neural Networks \and URM}

\section{Introduction}
The machine learning (ML) technique is known worldwide as a method to evaluate and predict the mechanical performance of civil engineering structures \citep{kouris2018state}. Neural networks (NN) have been successfully used to approximate the failure surface of such brittle anisotropic materials. 
 Previous studies have demonstrated that advanced metamodels, which employ full-field strain maps and loading scenarios, can accurately predict the mechanical properties of masonry elements. This approach can effectively reduce the need for expensive finite element simulations of masonry and can also support experimental testing in identifying masonry material properties both in situ and in the laboratory \citep{tahwia2021prediction,adaileh2023novel}. Other mechanical properties, such as the compressive strength of building elements, blast loading or detection of cracks, have been investigated with ML \citep{li2023determination,thango2023prediction}. In recent times, deep learning techniques have been employed to detect crack regions of historical masonry structures more precisely and sensitively using remote sensing technologies, with segmentation and object detection methods \citep{haciefendiouglu2023deep}. Machine learning (ML) algorithms and artificial neural networks (ANNs) have been developed to predict various properties of construction materials. For example, they can predict the shear strength of rectangular hollow reinforced concrete (RC) columns, the flexural response of steel-fiber-reinforced concrete (SFRC) beams, and the bond strength between masonry substrates and textile-reinforced mortar (TRM) strengthening system. Recent studies have reported that they can analyze the slump and uncover the internal dependencies within fiber-reinforced rubberized recycled aggregate concrete \citep{nguyen2023predicting,chained2024,soleymani2024textile,pal2024data,sun2020machine}. 
In brief, the applications in question underwent a review process and were subsequently classified into four distinct categories, namely: structural response and performance prediction, experimental data interpretation, image and text retrieval, and pattern recognition in structural health monitoring data \citep{srii2023prediction,zai2024damage,suwal2024plastic}. The accurate categorization of damages sustained by a building is a time-intensive process, with the premature analysis of the damage rate being crucial for mitigating the risks of potential accidents and the need for extensive repair work \citep{sri2023novel,ding2023improved}. Advanced algorithm models gave constructive insights into seismic design and evaluation of the cumulative damage of a reinforced concrete frame resulting from seismic sequences considering the structural performance of variable earthquake scale events \citep{lazaridisstructural,lazaridis2022applsci,lazaridis2023sustainability}. A comprehensive method has been developed for the mechanical characterization of TRM composites. This method uses the Finite Element Model to simulate various experimental tests on the TRM strengthening system, such as pull-out tests, tensile tests, shear bond tests, and in-plane shear tests. The results of these simulations demonstrate a safe approach to failure for the individual components, including the fibers' tensile failure, the mortar cracking and crushing, as well as the debonding strength of the fiber-mortar interfaces \citep{boem2022masonry}. 

Furthermore, ML valuably contributes to predicting the shear strength, the failure modes and health monitoring of masonry structures, which consist of a large part of cultural heritage buildings or masonry infill reinforced concrete frame structures, providing promising findings than typical processes \citep{plevris2014modeling,zhou2017shear,kaveh2023hybrid,thisovithan2023novel,loverdos2022automatic}. An area where this method can contribute in combination with smart health monitoring systems opens important perspectives in the future of existing constructions \cite{malekloo2022machine,naoum2023structural,chalioris2021flexural}. The combination of universal detection of critical information and monitoring, as well as the processing of large volumes of data, poses a challenge to research issues of using emerging technologies for structural integrity assessment.

To the best of the authors’ knowledge, there is no prior study that applies machine learning to predict the shear capacity of masonry walls with TRM jackets. To bridge this gap, the current study proposes a data-driven methodology that has been developed based on Machine Learning Methods to predict the shear capacity of unreinforced masonry walls strengthened with TRM strengthening system. More specifically, 17 ML algorithms were trained and tuned on experimental data selected from the bibliography, and their performances were compared to the test set. Subsequently, an evaluation of the two most capable regressors combined in a blended model to obtain even better performance. Subsequently, the rapid predictions of the ﬁnal model are explained and interpreted using ML interpretation methods. Finally, the conﬁgured methodology was implemented in a user-friendly graphical web application that is easy to access and employ for both researchers and engineers. 

\section{Feature Selection and Dataset Preprocessing}
In this section, the utilised input variables are described, and the overall pipeline through which the raw data passes before applying the ML algorithms is presented. The main key points of this preprocessing procedure are the feature scaling, one-hot encoding of categorical variables and the outlier detection-removal. The essential Python libraries used for the implementation of data preprocessing are the NumPy \citep{harris2020array} and Pandas \citep{mckinney2010data} for vectorized numerical calculations and dataset manipulation, respectively.

\subsection{Predictors of TRM-Strengthened URM Walls Shear Capacity}
The parameters studied in this research were chosen from the design relationships and the semi-empirical shear strength calculation relationship of Thomoglou et al. \citep{thomoglou2020ultimate,thomoglou2022experimental,thomoglou2023failure,thomoglou2023review}. Therefore, for the prediction of the shear capacity ($V_{exp}$) for three kinds of strengthened URM walls (brick, stone and cement), the significance of the following parameters is investigated. The parameter $A_n$ is the cross-sectional net area of the URM, the $f't$ is the tensile strength of the URM's binding mortar, the $t_f$ is the thickness of a network layer with fibers, the Ef is the Young modulus of the individual textile (grass, carbon, basalt), the Em is the Young modulus of the masonry wall, and the $f_c$ is the compressive strength of masonry. The $A_{mortar}$ is the area of the mortar of the strengthening system by unit width, and $E_{mortar}$ is the tensile modulus of elasticity of the cracked mortar of the TRM. The parameter $A_f$ is the area of the fabric reinforcement by unit width, the n is the number of layers of fabric, and finally, the $\varepsilon_{fu}$ is the ultimate tensile strain of the TRM reinforcement. The value ranges for each feature and corresponding publication are presented in Table~\ref{ranges}.

\subsection{One-Hot Encoding}
\begin{figure}
\includegraphics[width=8.8cm]{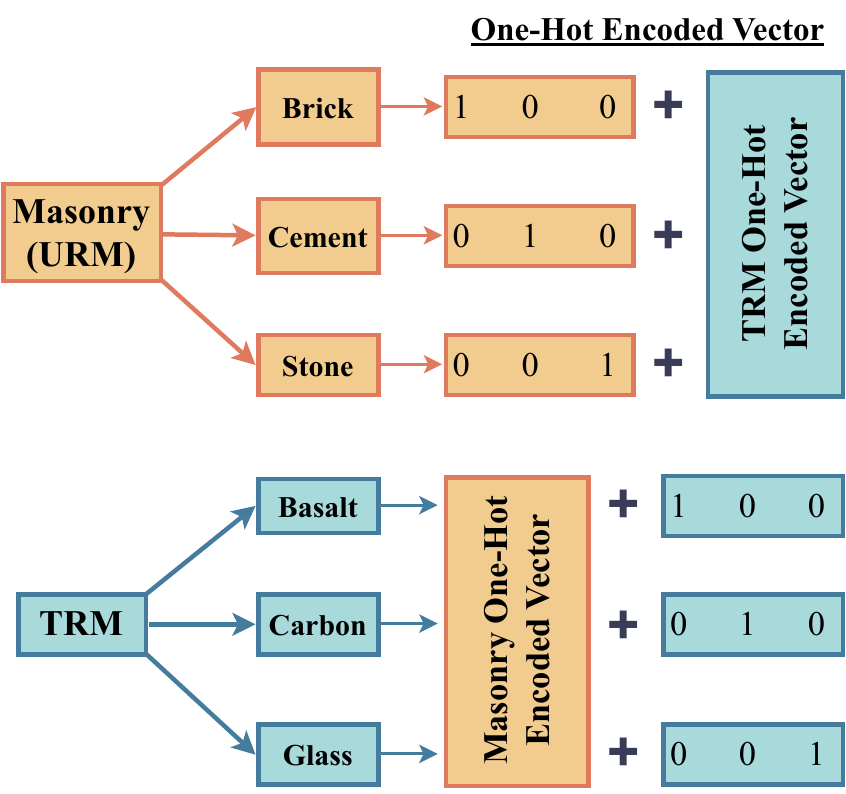}
\caption{One-Hot Encoding Visualization}\label{one_hot}
\end{figure}

The type of masonry and the type of TRM constitute the categorical input features of our regression problem. In the raw dataset, each category is represented by an alphabetical object as visualized in the second column of Fig.~\ref{one_hot}; to make them proper for the use of any ML algorithm, they need to be converted into a numerical format, according to one-hot encoding technique \citep{onehotencoding}. To achieve this, each categorical feature was divided into n binary (1 or 0) ones, representing the membership or not of each example in each of the n classes of the respective categorical variable; as a result, the two initial categorical predictors generated six binary ones. The type of masonry represented in the encoded dataset by the Brick, Cement and Stone attributes and the corresponding binary features of the TRM type are Carbon, Glass and Basalt. The number of examples in each category is visualised in Fig.~\ref{hists}a for the masonry type and in Fig.~\ref{hists}b for the TRM type. The generation of the one-hot encoded vectors for each data point is represented in Fig.~\ref{one_hot} in which the plus sign means vector concatenation.

\subsection{Feature Scaling}
\begin{figure*}
\includegraphics[width=\textwidth]{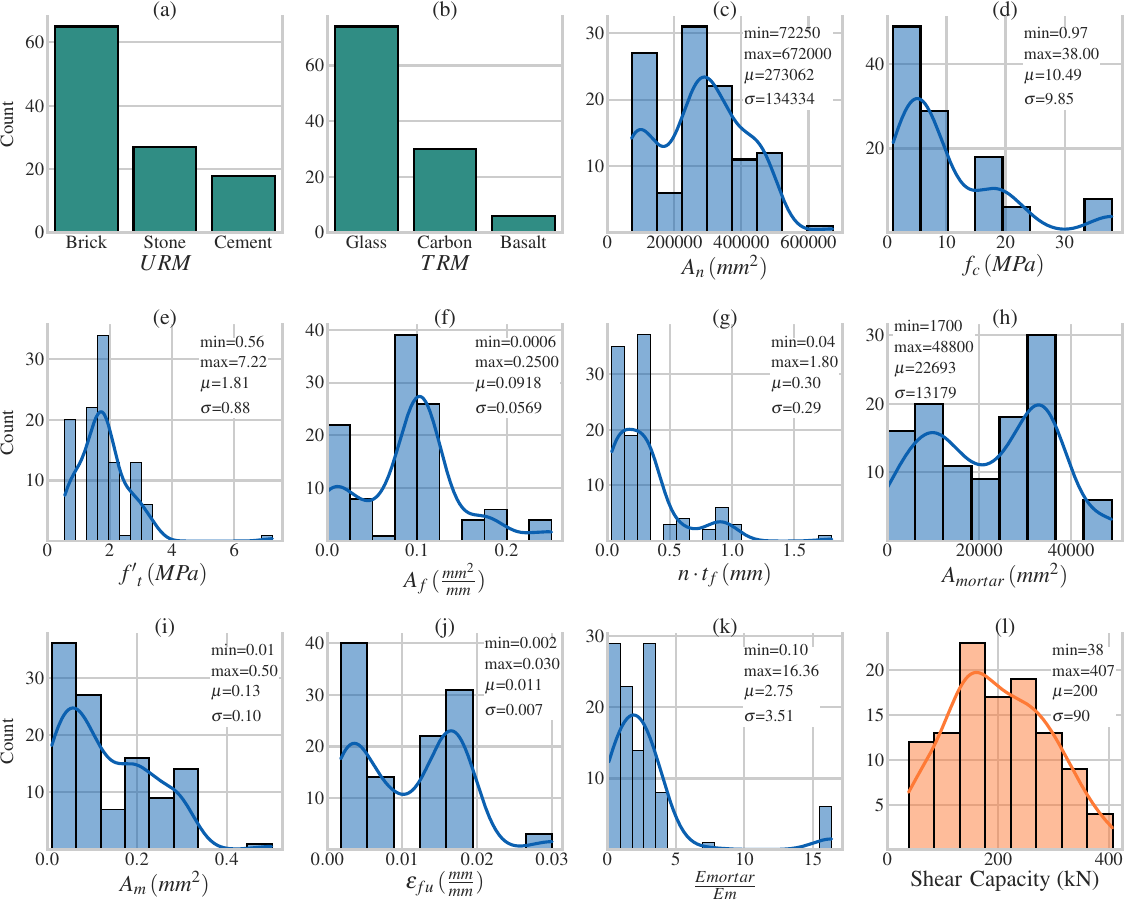}
\caption{(a,b) The bar plots of the categorical input variables, (c-k) the histograms of the numerical input features and (l) the histogram of the shear capacity}\label{hists}
\end{figure*}

In Fig.~\ref{hists}(c-k), it can be observed that our numerical input variables rely on different orders of magnitude. For instance, $A_n$ and $A_{mortar}$ are measured in $mm^2$, which are multiples of $10^4$ (Fig.~\ref{hists}h). Similarly, $\varepsilon_{fu}$ is measured in multiples of $10^{-2}$ (Fig.~\ref{hists}j), whereas $f'_t$ and $f_c$ are multiples of $10^0$ and $10^1$, respectively. In the context of ML methods such as Linear Models or Neural Networks, which take advantage of  Gradient Descent optimization algorithm variations, different magnitude orders of input features can cause convergence problems during training. To avoid this issue, the standard scaling \cite{muller2016introduction} was applied to each numerical predictor of our dataset. According to standard scaling, as presented in Equation~\ref{standard_scaling}, from the value of $m^{th}$ example and $k^{th}$ feature, the mean $\mu_{k}$ subtracted and the result divided with the standard deviation $\sigma_k$ of the corresponding input feature. The mean ($\mu$), the standard deviation ($\sigma$), the minimum and the maximum of each numerical input variable as well as for the target, could also be found in Figs.~\ref{hists}(c-k) and Fig.~\ref{hists}(l), respectively.

\begin{equation}
x_{mk}^{scaled} = \frac{x_{mk}-\mu_{k}}{\sigma_k}
\label{standard_scaling}
\end{equation}

\subsection{Outlier Detection}
\begin{figure}
\includegraphics[width=8.8cm]{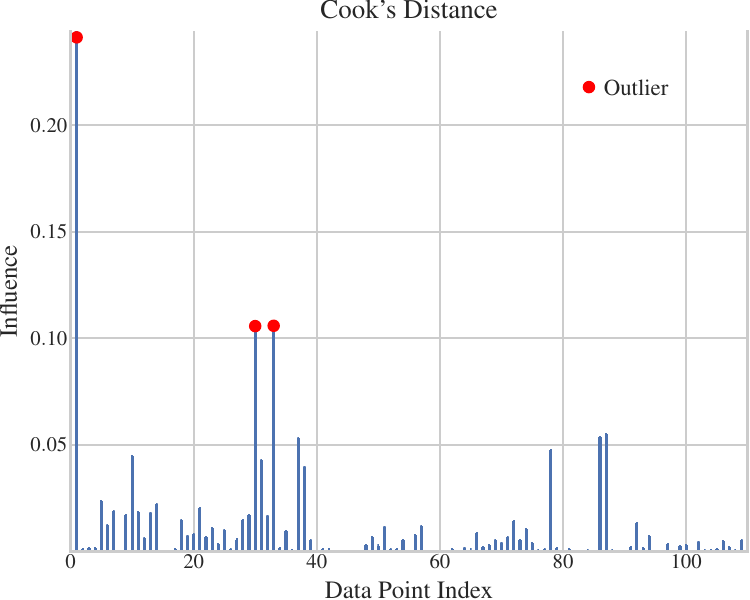}
\caption{Cook's Distance}
\label{cook}
\end{figure}

Observations that significantly vary from the majority of data can originate from various reasons, ranging from experimental errors to data entry mistakes known as outliers. To identify the possible outliers of the present dataset, the Cook's Distance \citep{cook1979influential} method was deployed. This process provides a measure for each data point, which indicates how influential this example is for a regression problem. A high value of Cook's Distance implies that the corresponding observation significantly impacts the model's coefficients. How much the removal of a data point i affects the regression coefficients is calculated as in Equation~\ref{cook_eq}. This equation calculates the sum of the square differences between the prediction which takes into account all observations ($\hat{y}_j$) and the corresponding without the $i^{th}$ observation ($\hat{y}_{j(i)}$), across all the (n) data points. This value is then divided by the product of the linear regression number of coefficients (p) and the mean squared error (MSE). The results of the Cook's Distance are presented in Fig.~\ref{cook} for each data point in our dataset after standard scaling and one-hot encoding. Three examples seem to have much more influence than the others, and for this reason removed from the dataset as outliers. The Cook's Distance of these points is represented with red markers in Fig.~\ref{cook}.

\begin{equation}
D_i = \sum_{j=1}^n\frac{(\hat{y}_j-\hat{y}_{j(i)})^2}{p\, MSE} 
\label{cook_eq}
\end{equation}

\begin{landscape}
\begin{footnotesize}
\begin{table}
\caption{The feature intervals and the number of experiments extracted for all publications from which the examined dataset mined.}\label{ranges}
\setlength{\tabcolsep}{4pt}
\footnotesize	
\renewcommand{\arraystretch}{1.1}
\begin{tabular}{lllllllllllr}
\toprule
Reference & $A_n\,({mm}^2)$ & $f_c\,(MPa)$ & ${f'}_t\,(MPa)$ & $A_f\,(\frac{{mm}^2}{{mm}})$ & $n\cdot t_f\,({mm})$ & $A_{mortar}\,({mm}^2)$ & $A_m\,({mm}^2)$ & $\varepsilon_{fu}\,(\frac{{mm}}{{mm}})$ & $\frac{Emortar}{Em}\,$ & $V_{exp}\,(kN)$ & Number Of Exp. \\
\midrule
Babadeidarabad \citep{babaeidarabad2014shear} & 72903-81290 & 19.0-22.0 & 2.21-3.28 & 0.05-0.10 & 0.05-0.19 & 11825-48800 & 0.01-0.01 & 0.01-0.01 & 0.46-0.48 & 108.8-246.7 & 12 \\
Borri et al. \citep{borri2014reinforcement} & 672000-672000 & 7.00-7.00 & 0.82-0.82 & 0.10-0.10 & 0.47-0.47 & 36000-36000 & 0.20-0.20 & 0.02-0.02 & 3.34-3.34 & 235.7-235.7 & 1 \\
Corradi et al. \citep{corradi2014shear} & 360000-360000 & 7.00-7.00 & 1.37-1.46 & 0.01-0.10 & 0.17-0.17 & 36000-36000 & 0.21-0.21 & 0.00-0.00 & 2.71-2.95 & 79.2-190.3 & 3 \\
Faella et al. \citep{faella2010shear} & 455900-480000 & 38.0-38.0 & 0.77-0.77 & 0.01-0.01 & 0.05-0.05 & 2355-4726 & 0.10-0.10 & 0.02-0.02 & 16.4-16.4 & 113.9-161.9 & 6 \\
Ferretti et al. \citep{ferretti2015situ} & 360000-360000 & 1.97-1.97 & 1.36-1.36 & 0.05-0.05 & 0.17-0.17 & 24000-24000 & 0.01-0.01 & 0.00-0.00 & 2.76-2.76 & 163.2-163.2 & 1 \\
Gams et al. \citep{gams2014seismic} & 312500-312500 & 0.97-0.97 & 1.70-1.70 & 0.03-0.03 & 0.47-0.47 & 2500-2500 & 0.01-0.01 & 0.03-0.03 & 1.78-1.78 & 157.3-157.3 & 1 \\
Gatesco et al. \citep{gattesco2015experimental} & 290000-464000 & 2.61-5.26 & 1.32-1.77 & 0.11-0.12 & 0.33-0.33 & 34800-34800 & 0.19-0.30 & 0.01-0.02 & 2.28-2.93 & 234.6-406.8 & 26 \\
Ismail et al. \citep{ismail2018plane} & 175560-232100 & 16.9-17.2 & 2.75-2.78 & 0.09-0.17 & 0.80-1.80 & 8360-18990 & 0.11-0.15 & 0.00-0.00 & 0.52-0.53 & 127.3-286.4 & 12 \\
Lignola et al. \citep{lignola2009nonlinear} & 257500-257500 & 5.00-5.00 & 0.76-0.76 & 0.09-0.18 & 0.04-0.08 & 16480-24720 & 0.05-0.05 & 0.02-0.02 & 1.45-1.45 & 80.4-122.5 & 4 \\
Mazzotti et al. \citep{mazzotti2016diagonal} & 360000-360000 & 8.05-8.05 & 2.07-2.07 & 0.05-0.06 & 0.17-0.17 & 24000-24000 & 0.01-0.01 & 0.00-0.00 & 1.19-1.19 & 296.1-331.2 & 2 \\
Mustafaraj \& Yardim \citep{mustafaraj2018plane} & 300000-300000 & 5.68-5.68 & 1.45-2.10 & 0.10-0.10 & 0.33-0.33 & 30000-30000 & 0.05-0.05 & 0.01-0.01 & 1.17-2.43 & 133.8-197.2 & 3 \\
Mustafaraj \citep{mustafaraj2016external} & 300000-300000 & 5.68-5.68 & 2.19-2.19 & 0.10-0.10 & 0.33-0.33 & 30000-30000 & 0.05-0.05 & 0.01-0.01 & 1.07-1.07 & 133.9-197.3 & 3 \\
Papanicolaou et al \citep{papanicolaou2007textile} & 72250-110200 & 3.90-8.20 & 1.39-1.67 & 0.01-0.25 & 0.05-0.09 & 1700-4640 & 0.07-0.11 & 0.00-0.01 & 0.80-2.66 & 38.3-58.8 & 8 \\
Parisi et al. \citep{parisi2013plane} & 381300-381300 & 2.50-2.50 & 2.40-2.68 & 0.10-0.10 & 0.17-0.17 & 12300-12300 & 0.01-0.01 & 0.02-0.02 & 0.31-0.39 & 263.1-328.0 & 2 \\
Prota et al. \citep{prota2006experimental} & 257500-257500 & 5.00-5.00 & 0.76-0.76 & 0.09-0.18 & 0.04-0.08 & 10300-20600 & 0.05-0.05 & 0.02-0.02 & 3.86-3.86 & 88.0-165.7 & 8 \\
Shabdin et al. \citep{shabdin2018experimental} & 120000-120000 & 8.80-8.80 & 1.82-1.82 & 0.00-0.00 & 0.30-0.60 & 18000-30000 & 0.08-0.08 & 0.00-0.00 & 1.54-1.54 & 62.9-251.0 & 7 \\
Thomoglou et al. \citep{thomoglou2022experimental} & 249900-357000 & 5.3-38.0 & 1.32-7.22 & 0.10-0.10 & 0.09-0.51 & 11900-30000 & 0.16-0.23 & 0.00-0.00 & 0.10-2.95 & 169.7-328.9 & 9 \\
Tomazevic et al. \citep{tomavzevivc2015strengthening} & 325625-375000 & 1.14-1.14 & 0.56-1.36 & 0.24-0.24 & 0.17-0.17 & 25000-26050 & 0.26-0.50 & 0.03-0.03 & 2.17-7.05 & 176.9-204.2 & 2 \\
\bottomrule
\end{tabular}

\end{table}
\end{footnotesize}
\end{landscape}

\section{Machine Learning: Application, Cross-Validation, Interpretation and Deployment}
\begin{figure*}
\includegraphics[width=\textwidth]{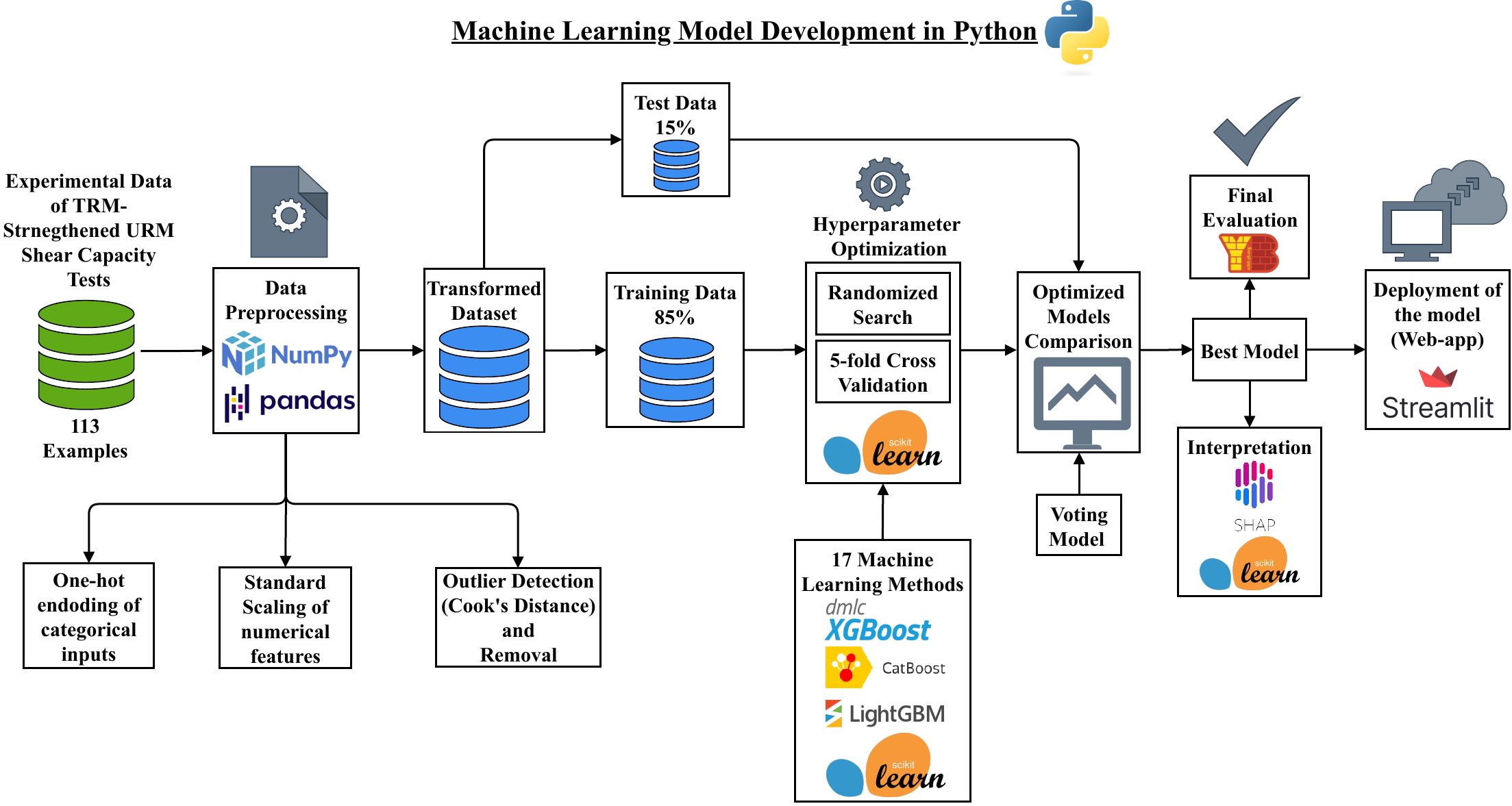}
\caption{The implemented development workflow of the proposed model using the Python Machine Learning Ecosystem.}\label{flowchart}
\end{figure*}
\begin{figure*}
\includegraphics[width=\textwidth]{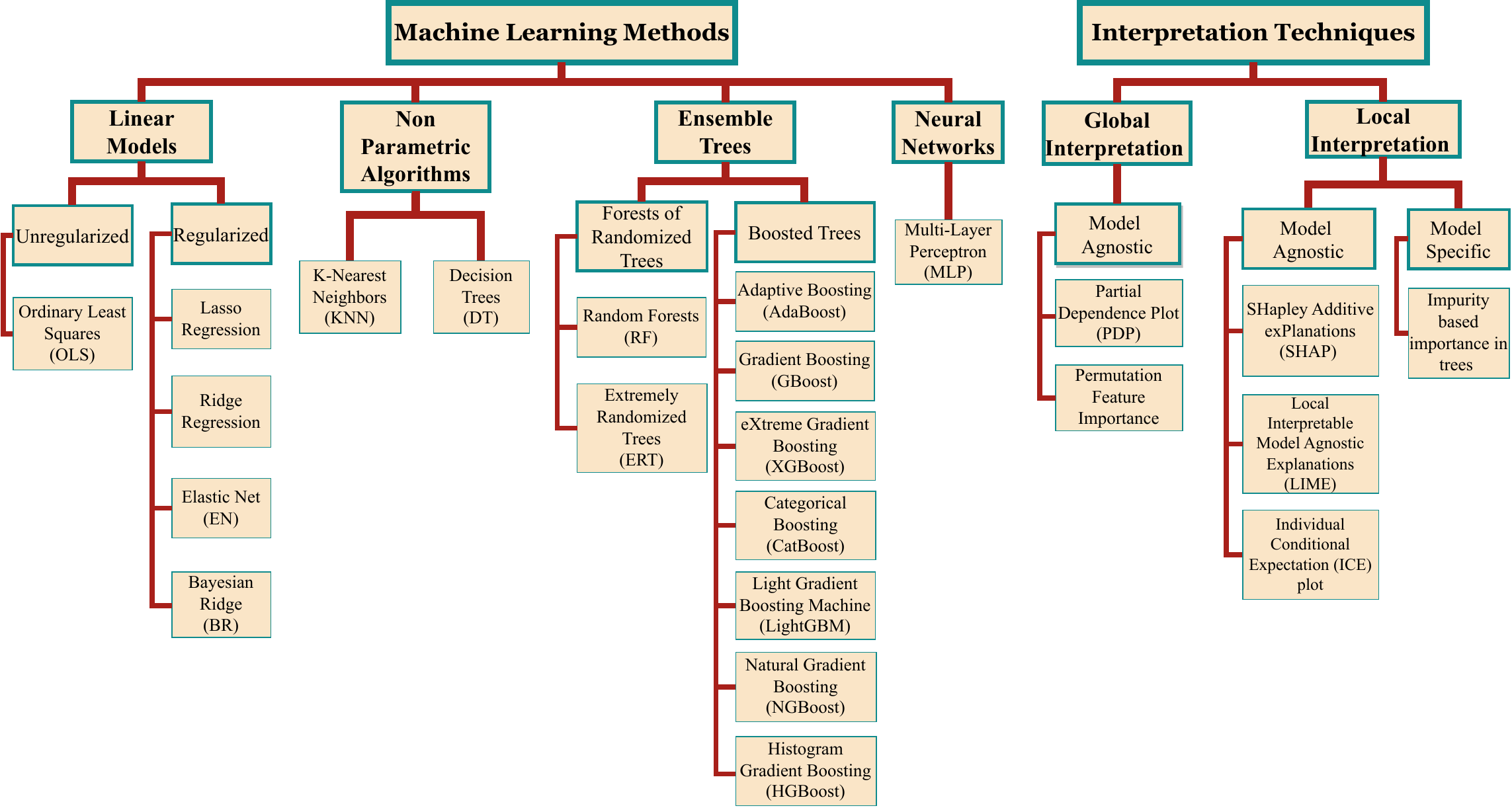}
\caption{A comprehensive representation of the examined Machine Learning Methods and Interpretation Techiniques}\label{models_org}
\end{figure*}

To develop an effective and robust ML model for predicting TRM-strengthened URM shear capacity, it is crucial to investigate several methods, from basic and simple algorithms to the most recently developed and complex techniques. For this purpose, 17 ML regression methods examined in this study consisted of Linear Models, Non-parametric algorithms, Ensemble Trees and Neural Networks (Fig.~\ref{models_org}). These methods were selected from the literature as they are suitable for analyzing a tabular dataset like the one being analysed in the current study. More generally, ML methods aim to learn from data and make reliable predictions for new unseen cases. To target that, the total dataset is split into two primary subsets: the training and the test set. The former serves for the ML algorithm fitting, while the latter is used for the unbiased evaluation of the trained model performance. Machine learning models optimize the cost function during training to reduce the total error between actual and predicted values. If the complexity of the model is higher than required, then the model is sensitive to the overfitting problem, which leads to much better performance in training than this of the test set. To control the complexity, many models incorporate a regularization hyperparameter in their cost function, which penalizes the unnecessary increase of the model parameters. Suppose more than the complexity of the model is needed to describe the mapping between input and output variables. In that case, the model suffers from high bias or underfitting, leading to low performance in both training and test sets. In our study, the overall dataset after the preprocessing consists of 110 examples. To ensure adequate data for both training and testing of ML algorithms, the dataset was randomly split into 85\% for training and 15\% for testing. After preprocessing, our dataset includes six binary input features and 11 numerical predictors, a total of 17 input variables and one target. The overall workflow adopted in this study, as well as the corresponding Python libraries employed at each step, is visualized in Figure~\ref{flowchart}. In this section, the applied ML regression and interpretation methods are described, and the corresponding results are presented and discussed.

\begin{equation}
\mathcal{J} = \frac{1}{m}\sum_{k=1}^m\mathbcal{L}\left(y_k, \hat{y}_k\right) + regularization\,\,terms
\label{cost_function}
\end{equation}

\subsection{Linear Models}
Linear models are the most fundamental family of regression methods, which includes the Ordinary Least Squares (OLS), the Lasso \citep{tibshirani1996regression}, Ridge \citep{hoerl1970ridge}, Elastic Net \citep{zou2005regularization} and the Bayesian Ridge \citep{scikit-learn} regression methods. The Equation~\ref{linear_cost} demonstrate the general cost function of a regularised linear model. The first term corresponds only to the OLS method, while the addition of the second configures the cost function of the Lasso regression. More concretely, the norm-1 of the model coefficients is multiplied by the regularization parameter ($\alpha\rho$) and added to the OLS cost function. Through this kind of regularization, the weights of the most non-informative predictors become zero. The summation of the first and the third terms makes the cost function of the Ridge regression, which penalizes the increasing of the model weights using the norm-2 and mandates the weights of non-important features to take smaller values. Finally, the overall equation represents the cost function of the Elastic Net method, which is the superposition of both regularization ways. 
\begin{equation}
\hat{y}\left(x_j\right)  \,=  \,b + \sum_{j=1}^nw_jx_j\label{linear_comb}
\end{equation}

\begin{equation}\label{linear_cost}
\begin{split}
\mathcal{J}(w_j, b)  =   \frac{1}{2m}\underbrace{{\sum}_{k=1}^m\left(\hat{y}(x_{jk}) - y_k\right)^2}_{OLS} + \,\alpha\rho\underbrace{\sum_{j=1}^n \left|w_j\right|}_{\mathbcal{l}1} \\ +  \frac{\alpha(1-\rho)}{2}\underbrace{\sum_{j=1}^n w_j^2}_{\mathbcal{l}2}
\end{split}
\end{equation}
\subsection{Non-Parametric algorithms}
Non-parametric algorithms do not assume trainable parameters or weights to make predictions; instead, they adjust their structure based on the training data and the initial guess of their hyperparameters. In this study, the K-Nearest Neighbours (KNN) \citep{altman1992introduction} and the Decision Trees \citep{breiman2017classification} are studied as representative examples of this family of ML algorithms. In the KNN method, the efficiency of regression highly depends on the number of neighbours (K), which is the most crucial hyperparameter of the method. While the number of neighbours is reduced, the prediction is more susceptible to the errors of individual data points; also, if the K is too high, the model does not effectively utilize the different data patterns. In the Decision Trees, the prediction is made after a sequence of binary decisions at the nodes of the tree until a leaf is met, which gives the predicted value. During training, the tree selects the input feature at each node in which the spit is applied and the corresponding threshold to maximize the impurity reduction of the produced subsets. As the depth increases, the complexity of the tree increases, and the final prediction becomes more prone to overfitting.
\subsection{Ensemble Trees}
The ensemble trees are divided into two branches: Random Trees and Boosted Trees. Both families rely on combining the predictions of multiple learners to enhance the robustness and accuracy of estimation. The Random Trees varied in two versions, which were based on the voting of a population of independent models. The Random Forest \citep{breiman2001random} variation incorporates the prediction of multiple randomized Decision Trees. More concretely, each tree is fitted on a random subset of the training set, which was created using sampling with replacement. Also, the candidate features for splitting at each node is a random subset of the total available predictors. The use of many trees by this method and the introduction of randomness in the construction of the base trees seek to reduce the risk of overfitting stemming from the predictions of a single Decision Tree. The Extremely Randomized Trees (ERTs) \cite{geurts2006extremely} version also adds another source of randomness to the generation of the base trees. Simultaneously with the random choice of candidate split features at each node, the splitting threshold is also determined randomly for each candidate feature during the training of each single tree. The decoupled construction of the base trees in the random trees concept could be utilized from parallel computing resources to reduce the training time efficiently.

Contrarily, in the Boosted Trees method, multiple weak trees are generated sequentially instead of simultaneously and finally combined to form the ensemble model. The core idea is to gradually improve the model's accuracy by focusing on the parts of the dataset where the current model fails. The procedure starts with an initial tree model, which is trained on the examined dataset. This initial model is usually fundamental and provides a starting point. After this stage, the algorithm evaluates the model's performance and identifies the data points where the more significant errors are made. A new weak model is trained to reduce mainly the errors of the above data points and added to the previous one after the multiplication with the learning rate ($\gamma$). This learning rate is responsible for determining how fast the overall model is fitted on the training set. A high value of this coefficient could lead to overfitting; conversely, a learning rate that is too low can lead to an inefficient and slow learning process. The Adaboost \citep{freund1997adaboost,drucker1997improving} algorithm is the older variation of the boosting method and is the most prone to overfitting. Gradient Boosting (GBoost) \citep{friedman2001greedy} was a pioneer ML technique which combines the boosting meta-algorithm with the Gradient Descend optimizer to minimize the errors between predictions and actual values of the target variable. Since Gradient Boosting was introduced, different versions of the method with several improvements have been proposed. XGBoost (eXtreme Gradient Boosting) \citep{chen2016xgboost} is an evolution of the traditional gradient boosting algorithm and deals with the efficient regularization, scalability, and portability of its application. Histogram-based Gradient Boosting Regression Tree (HGBoost) \citep{scikit-learn} discretizes the feature space into bins to accelerate the process of estimating the optimal splits. On the other hand, CatBoost \citep{dorogush2018catboost,prokhorenkova2018catboost} specializes in dealing with categorical variables by implementing advanced strategies such as ordered boosting and oblivious trees. Additionally, the Natural Gradient Boosting (NGBoost) algorithm \citep{ngboost}, which was developed by Stanford University researchers, introduces the uncertainty of the prediction using probabilistic modelling. Finally, LightGBM \citep{ke2017lightgbm} was created by Microsoft and employs an optimized histogram decision tree algorithm that focuses on improving resource usage and training speed. 

The Gradient Boosting equation is expressed as:

\begin{equation}\label{voting_gb_cost}
\mathcal{J}(\beta_n)= \sum_{i=1}^{m} \mathcal{L}(y_i, F_{m-1}(x) + \gamma_m h_m(x))
\end{equation}
where
$F(x)$ is the final prediction of the model,
$F_{n-1}(x)$ is the prediction of the previous iteration,
$T_n(x)$ is the n-th tree model in the ensemble,
$\beta_n$ is the weight assigned to the n-th tree model, 
and $m$ is the number of samples.

\begin{equation}
F_m(x) = F_{m-1}(x) + \gamma_m h_m(x)
\end{equation}

This equation encapsulates the iterative update process in gradient boosting. Let's break down each term:

\begin{itemize}
    \item $F_m(x)$: This represents the predictive model at the $m^{th}$ iteration. As the iterations progress, $F_m(x)$ becomes an increasingly refined version of the model, improving its predictive accuracy by incorporating the learnings from previous iterations.
    \item $F_{m-1}(x)$: This term is the model as it stood in the previous iteration, $(m-1)^{th}$. It is the foundation for the new model $F_m(x)$.
    \item $\gamma_m$: Known as the learning rate, this term scales the contribution of the newly added model, $h_m(x)$. It is a crucial parameter that balances the learning speed against the risk of overfitting. A smaller $\gamma_m$ leads to slower, more cautious model updates, while a larger $\gamma_m$ makes the model adapt more quickly, potentially at the risk of overfitting to the training data.
    \item $h_m(x)$: This is the new model trained during the $m^{th}$ iteration, specifically targeting the residual errors made by $F_{m-1}(x)$. By focusing on these errors, $h_m(x)$ effectively corrects the mistakes of the previous model, refining the overall predictive power of $F_m(x)$.
\end{itemize}

Through each iteration, the model's performance is gradually enhanced, with the goal of minimizing prediction errors on the training dataset. This iterative refinement makes gradient boosting a powerful tool for complex regression and classification tasks.

\subsection{Neural Networks}
\begin{figure}
\includegraphics[width=8.8cm]{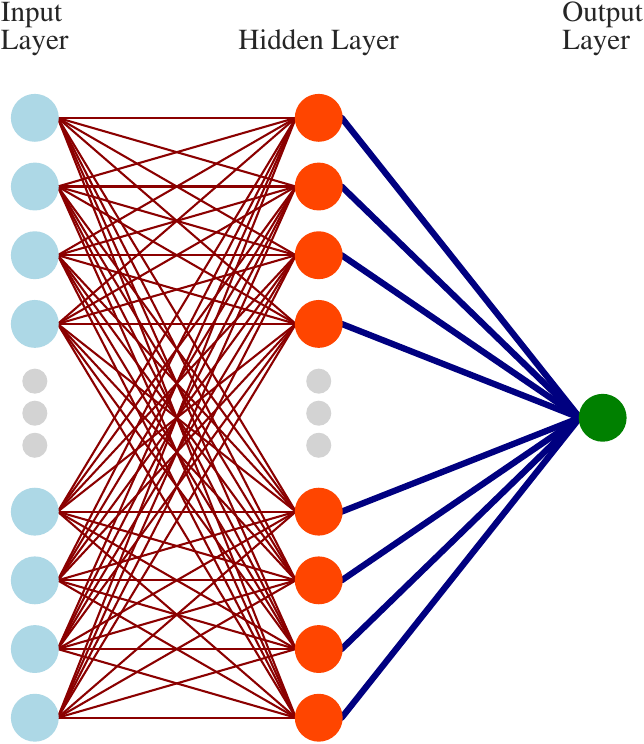}
\caption{A shallow Feedforward Neural Network Visualization}\label{mlp_viz}
\end{figure}

 The Multi-Layer Perceptron (MLP) is a type of neural network that is suitable for tabular data, as the case in the presented study. The MLP is comprised of successive dense and fully connected layers as visualised in Figure~\ref{mlp_viz}. The figure illustrates a shallow neural network with a single hidden layer and a variable number of hidden neurons, as examined in the present paper. After training, MLPs can capture complex mappings by taking advantage of the nonlinear activation function applied at each hidden neuron. The Forward Pass (Equation~\ref{mlp_for}) calculates the activation value $z_i^\mathbcal{l}$ of the neuron i and applies the activation function ($\sigma$) on a linear combination of the previous layer ($\mathbcal{l}-1$) neuron activations. In sequence, during training, the Backward Pass (Equation ~\ref{bp}) \cite{rumelhart1986learning,lecun2012efficient} calculates the derivative of the cost function $\mathcal{J}$ with respect to the model weights ($W_{ij}^\mathbcal{l}$) and biases ($b_i^\mathbcal{l}$), known as trainable parameters. Both forward and backward passes are required to implement one iteration of the Gradient Descent optimization algorithm. The training dataset is divided into a number of batches. At each step of the Gradient Descent (Equation~\ref{gd}), an update of the model trainable parameters is implemented using one batch of the training dataset. After every batch has undergone the optimization process, an epoch is completed once the entire training dataset has gone through the process. Multiple epochs are usually demanded by the cost function (Equation~\ref{mlp_cost}) to reach a minimum, and therefore the examined MLP to fit on the training dataset. Once the MLP is trained, only one iteration of the Forward Pass is needed to predict the target value for a new input.
\begin{equation}\label{mlp_for}
a_i^\mathbcal{l} = \sigma\underbrace{\left(\sum_{j}W_{ij}^\mathbcal{l}a_j^{\mathbcal{l}-1} +b_i^\mathbcal{l}\right)}_\text{$z_i^\mathbcal{l}$}
\end{equation}
where $L$ is the number of layers, n is the number of neurons of layer $\mathbcal{l}$, m is the number of neurons of layer $\mathbcal{l}-1$.

\begin{equation}\label{bp}
\frac{\partial \mathcal{J}}{\partial W_{ij}^\mathbcal{l}} = \left(\frac{\partial \mathcal{J}}{\partial a_i^\mathbcal{l}} \odot \frac{\partial a_i^\mathbcal{l}}{\partial z_i^\mathbcal{l}}\right)a_j^\mathbcal{l-1}
,\,\,\,\,\,\,\,
\frac{\partial \mathcal{J}}{\partial b_i^\mathbcal{l}} = \frac{\partial \mathcal{J}}{\partial a_i^\mathbcal{l}} \odot \frac{\partial a_i^\mathbcal{l}}{\partial z_i^\mathbcal{l}}
\end{equation}
where $\odot$ denotes element-wise multiplication. 

\begin{equation}\label{gd}
W_{ij}^\mathbcal{l} \leftarrow W_{ij}^\mathbcal{l} - \alpha \frac{\partial \mathcal{J}}{\partial W_{ij}^\mathbcal{l}}
,\,\,\,\,\,\,\,
b_i^\mathbcal{l} \leftarrow b_i^\mathbcal{l} - \alpha \frac{\partial \mathcal{J}}{\partial b_i^\mathbcal{l}}
\end{equation}
where $\alpha$ is the learning rate, and $\leftarrow$ denotes assignment.

\begin{equation}
\mathcal{J}{reg}(W_{ij}, b_i) = \mathcal{J}(W_{ij}, b_i) + \lambda \sum_{\mathbcal{l}=1}^{L} \sum_{i=1}^n \sum_{j=1}^m {\left(W_{ij}^\mathbcal{l} \right )}^2
\label{mlp_cost}
\end{equation}
where $\lambda$ is the regularization parameter

\subsection{Hyperparameter Tuning and Cross Validation}
\begin{figure}
\includegraphics[width=8.8cm]{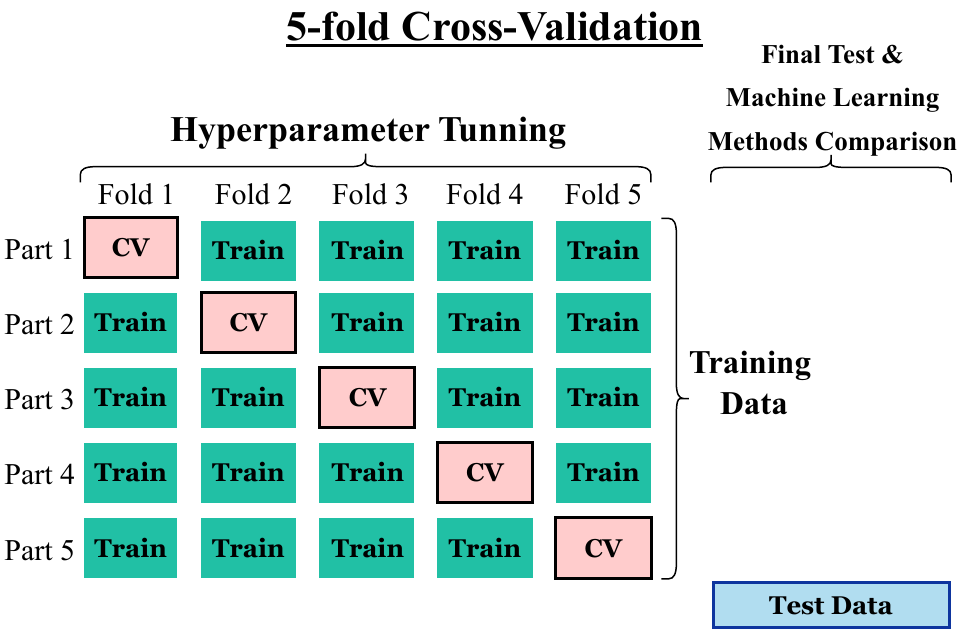}
\caption{The k-fold cross validation strategy}\label{k_fold}
\end{figure}

In contrast with trainable model parameters, the hyperparameters must defined before the model training. Usually, these parameters control the complexity of the model, such as the maximum depth in a decision tree-based model, the number of hidden layers and neurons in the case of feedforward neural networks or the regularization factor, which several methods include in their cost function. Additionally, some of these parameters affect the model's assumptions regarding the data or the training process, such as the activation function and the learning rate respectively. In our study, the  Randomized Search Cross-Validation (CV) \citep{bergstra2012random} was used to select the candidate hyperparameter values for each investigated ML regression method. In the Linear model, the regularization parameter values are tested within the range of 0 to 10. Meanwhile, for KNN, the number of neighbours is considered to be between 5 and 20. In the case of decision trees, two hyperparameters are investigated: the maximum depth of the tree within the range of 2 to 10 and the splitting criterion between Squared Error, Friedman MSE, Absolute Error, and Poisson. Further, in the case of Boosted Trees, the learning rate is considered a hyperparameter with possible values from 0.01 to 1. The Randomized trees are tested with two hyperparameters, the number of base trees and the maximum tree depth in the intervals [10, 200] and [2, 10], respectively. In the case of Multi-Layer Perceptron (MLP) Neural Networks, \cite{rojas1996neural} as hyperparameters included: the number of neurons, the regularization parameter and the activation function of the hidden neuron among Sigmoid, ReLU (Rectified Linear Unit), Tanh and Identity. Considering the possible intervals and sets of hyperparameter values defined above for each ML method family, using the Randomized Search CV selected 100 instances for every regression technique. Each candidate model is trained and evaluated according to a k-fold cross-validation strategy \citep{picard1984cross} with k equal to 5.  According to the k-fold CV technique, the training dataset split randomly at k parts as visualised in Figure~\ref{k_fold}. Each fold permutes a different part as the test and the remaining parts as the training set. The performance of the model is evaluated for each fold, and the best model is selected based on the mean performance across all five folds. All the above training and evaluation procedures were implemented using scikit-learn \citep{scikit-learn} and Yellowbrick \citep{bengfort_yellowbrick_2019} Python frameworks.

\subsection{Comparative Analysis of Machine Learning Algorithms}
\begin{figure*}
\includegraphics[width=\textwidth]{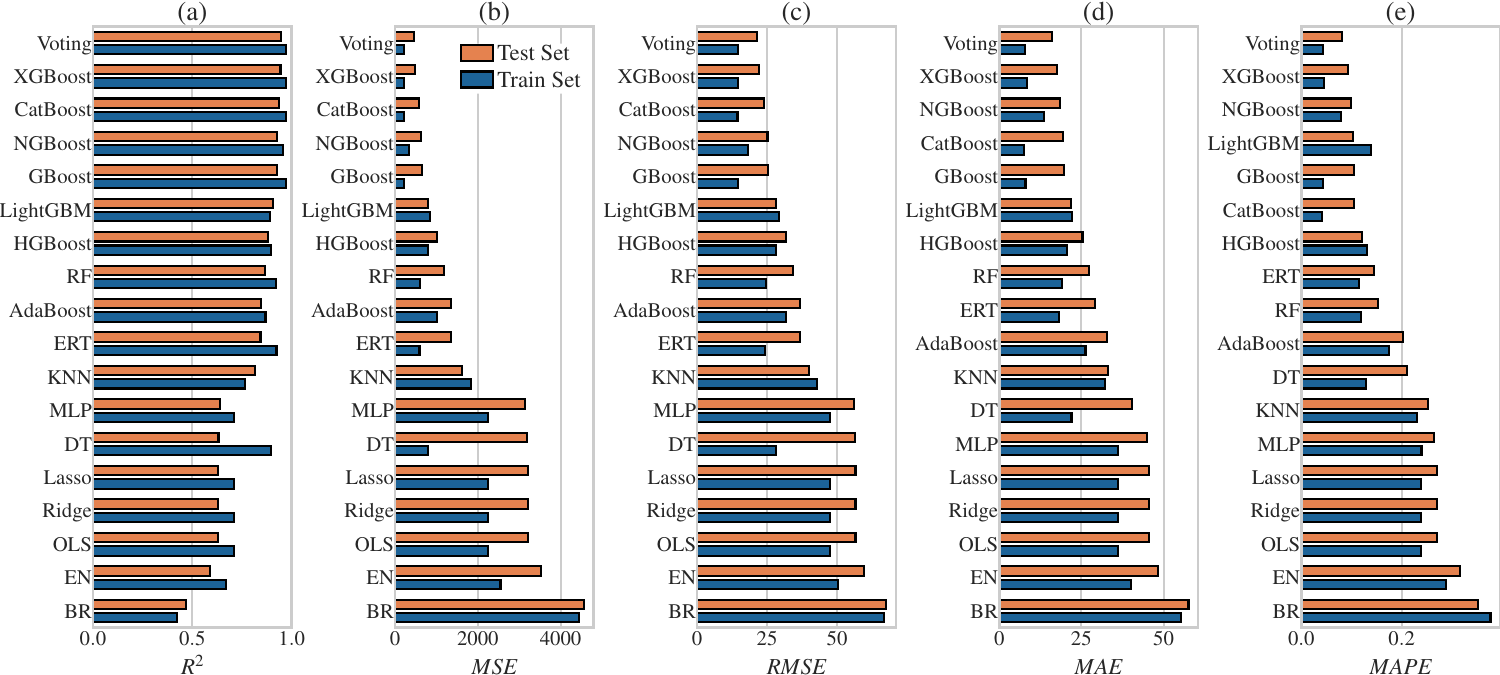}
\caption{Comparison of the examined Machine Learning methods performances according to (a) $R^2$, (b) MSE, (c) RMSE, (d) MAE, (e) MAPE}\label{model_comp}
\end{figure*}

The ability of the examined ML algorithms to model our regression problem was assessed using five different evaluation metrics. In Fig.~\ref{model_comp}, the implemented ML methods are sorted in descending order of effectiveness on the test set for the five evaluation metrics. In case $R^2$ descending order and in ascending for the error metrics. We can observe that the Decision Tree model fails to generalize its predictions in the test set and succumbs to overfitting due to the large gap between training and test performances. The prediction ability of this model is excellent on the training set but much worse on the test set. The MLP regressor with a single hidden layer with 80 neurons, activated with ReLU and regularization rate of 3.8 and batch size of 32 examples, exhibits the best performance of the examined MLPs configurations with $R^2$ equal to 0.79 and 0.71 on the training and test set respectively. The ensemble trees present the best prediction capacity among all the ML method families, which are investigated in this study as shown in Fig.~\ref{model_comp}. An XGBoost model, with a maximum depth of 5 and a learning rate of 0.12, exhibits the best prediction performance on the test set with an $R^2$ value of 0.94 and MAPE of 9.3\%. The CatBoost model, with the same maximum depth and a learning rate of 0.2, displays the second best prediction ability in terms of 3 out of 5 performance metrics, with an $R^2$ value of 0.93 and a MAPE of 10.5\%, on the test set. To configure an even more unbiased model, the combination of the two most capable methods (XGBoost, CatBoost) is implemented in a Voting regressor, which averages their predictions. The result blended model exhibits improved performance in comparison with all the other models (Fig.~\ref{model_comp}) with an $R^2$ equal to 0.97 and 0.95 on training and test sets, respectively. Also, all its error metrics are reduced compared to individual models with MAPE equal to 4.33\% and 8.03\% for training and test sets, respectively. The performance metrics that have been applied are presented below.
\begin{enumerate}
\item The coefficient of determination, $R^2$, expresses the variation in the dependent variable that is predictable from the independent variables:
\begin{equation}
R^2 = 1 - \frac{\sum^n_{i=1}(y_i-\hat{y}_i)^2}{\sum^n_{i=1}(y_i-\overline{y}_i)^2}
\end{equation}

where $\overline{y}$ is the average of the observed values;

where  $\hat{y}_i$ is the predicted value, $y_i$ the real value of the $i${th} observation and n is the total number of observations;
\item Mean square error ({MSE}):
\begin{equation}
MSE = \frac{1}{n}\sum^n_{i=1}(\hat{y}_i-y_i)^2
\end{equation}

\item Root-mean-squared error ({RMSE}) calculates the average error between the estimated values and the observed values:
\begin{equation}
RMSE = \sqrt{\frac{1}{n}\sum^n_{i=1}(\hat{y}_i-y_i)^2}
\end{equation}

\item Mean absolute error ({MAE)}
 is a measure of errors between the estimated and the observed values, and it is given by the following expression:
\begin{equation}
MAE = \frac{1}{n}\sum^n_{i=1}\mid\hat{y}_i-y_i\mid
\end{equation}

\item Mean absolute percentage error ({MAPE}) calculates the accuracy as a ratio and is defined by the following formulation:
\begin{equation}
MAPE = \frac{1}{n}\sum_{i=1}^n\left|\frac{y_i-\hat{y}_i}{y_i}\right|
\end{equation}
\end{enumerate}

\subsection{Evaluation of the Best Model}
\begin{figure}
\includegraphics[width=8.8cm]{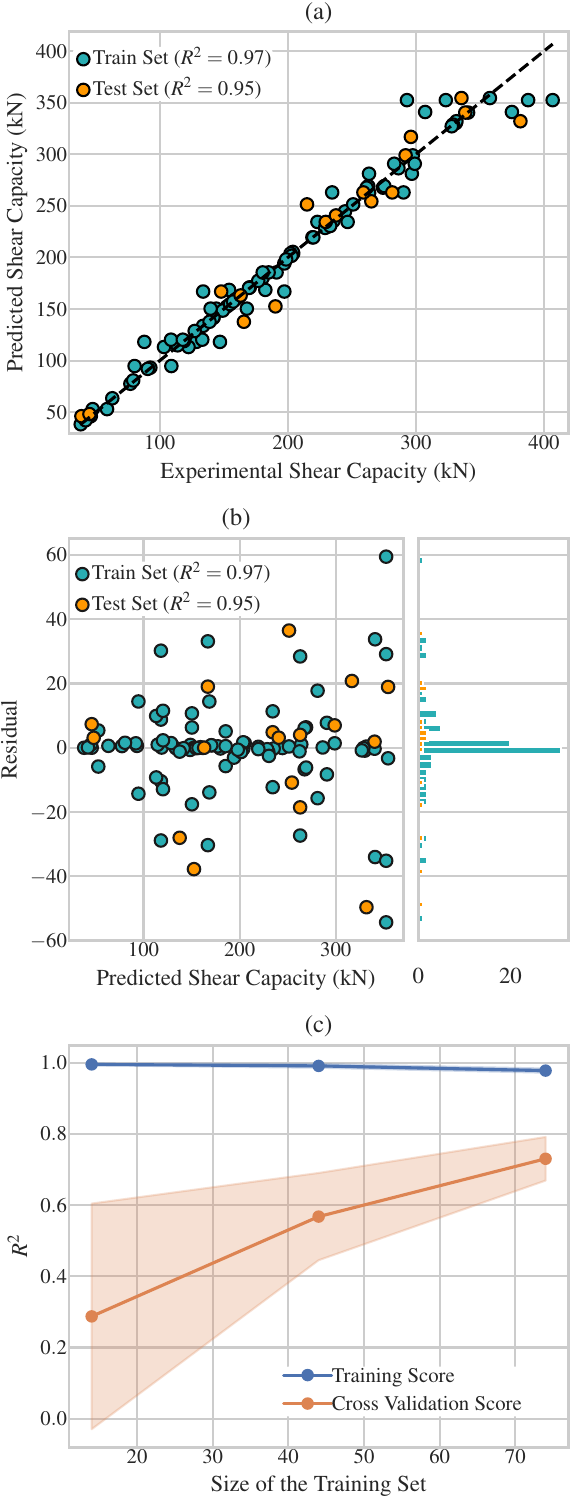}
\caption{Evaluation of the Voting model using (a) Actual vs Predicted plot (b) Residual plot and (c) the Traning Curve}\label{model_eval}
\end{figure}

Fig.~\ref{model_eval}a depicts the experimental against the predicted TRM-strengthened masonry shear capacity values for both the train and the test set with green and orange points, respectively. The prediction capacity of the configured Voting model in terms of determination coefficient is $R^2=0.97$ for the train set and $R^2=0.95$  for the test set. The black dashed line denotes the identity function and helps us assess how well the model prediction agrees with the corresponding experimental values. Fig.~\ref{model_eval}b illustrates the predicted values of the shear capacity versus the differences between the actual and predicted values both in the train and test set, which are called residuals. Also, in the right place of the same figure, the respective histograms are presented for both sets. This figure helps to recognize if any patterns exist in the scatter plot of residuals and validate our regression behaviour. There is no observed pattern, and the residuals are randomly scattered with an approximately symmetrical distribution around zero. Additionally, the errors are more densely distributed near zero and less so as they move away from it. Fig.~\ref{model_eval}c illustrates the learning curve of the proposed model. The x-axis denotes the size of the training set the model fitted on at each point of the training curve, while the y-axis indicates the determination factor. The two lines represent the effectiveness of the proposed model as the training set increased. The blue line is for the training set, and the orange line is for the validation set. At each step of the learning curve, the size of the training set is constant, but according to the 5-fold cross-validation, a permutation occurs between the training and validation set as explained in Fig~\ref{k_fold}. As an outcome of the above, the shadows in Fig.~\ref{model_eval}c display the uncertainty sourced from the different training and test selections of the k-fold cross-validation process. At the same time, the points represent the mean determination factor at each step of the learning curve. At the starting point, the model seems to have a cross-validation $R^2$, approximately equal to 0.3 and a high variance problem. As the size of the training set increased, the cross-validation score was raised, and its uncertainty was decreased, making the model more robust to different train-test set splittings. Also, the training and cross-validation scores converged to each other, which led to the reduction of overfitting and the improvement of the model generalization ability. The above behaviour reveals that the model has the potential to learn even better as the size of the training set increases, which gives us more confidence about the final deployment where the model will be fitted not only on the training set but on the overall dataset, as usual in ML practice.

\subsection{Comparison with existing models}
\begin{figure*}
\includegraphics[width=\textwidth]{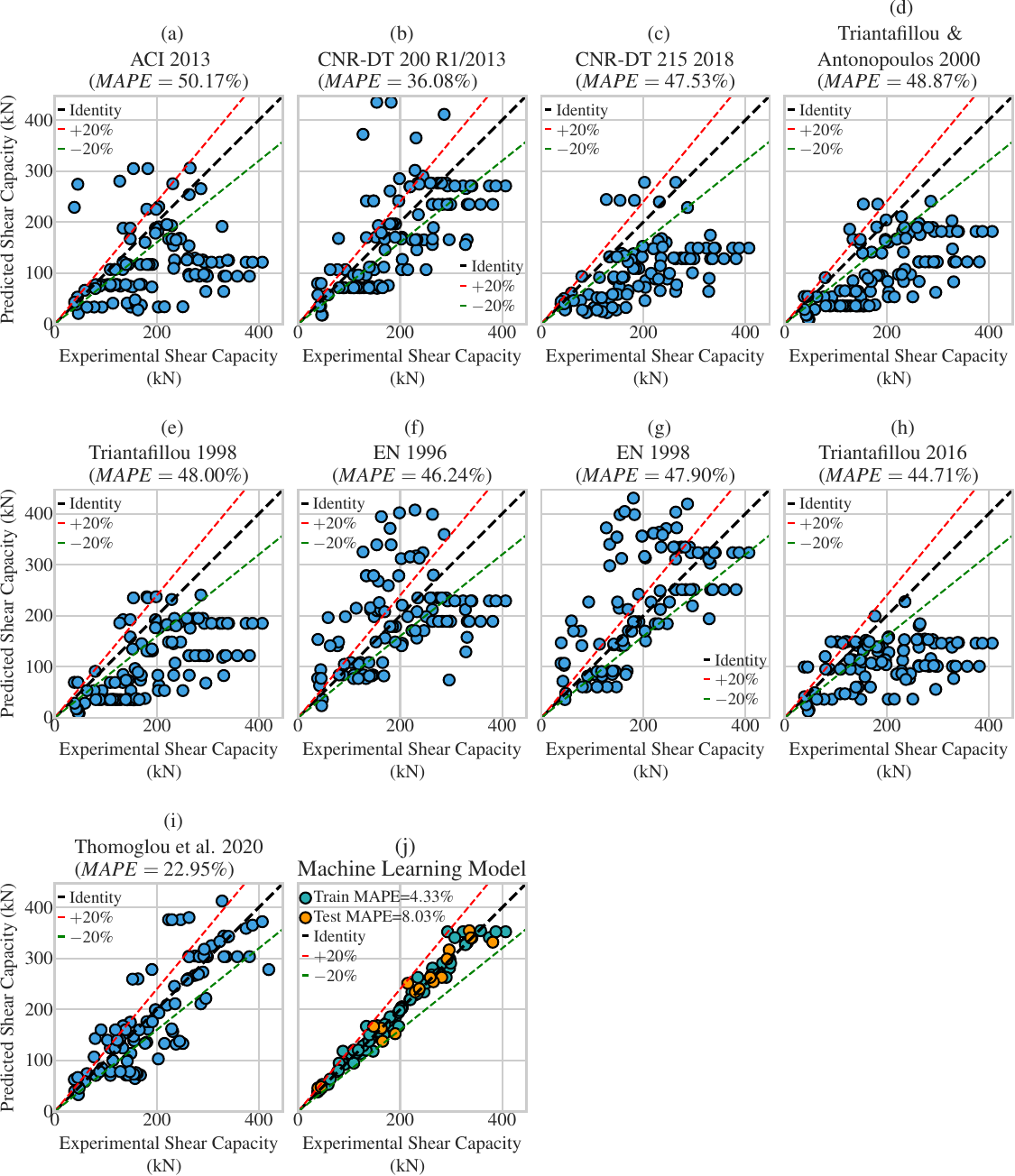}
\caption{Comparison of the proposed Machine Learning model with existing models.}\label{comp_previous}
\end{figure*}

In Fig.~\ref{comp_previous},  the performance on the examined dataset for a series of existing models as well as for the proposed model is visualized. The scatter plots of Fig.~\ref{comp_previous}(a-i) illustrate the experimental versus predicted shear capacity of analytical models that have been proposed by several researchers and codes in the past. All subfigures of Fig.~\ref{comp_previous} share both the x and y axes, making it easy to compare different models directly. The black dashed line represents the identity function, which is the ideal geometric place for the predictions of each model. Also, the red and green dashed lines correspondingly denote $20\%$ overestimation or underestimation of the predicted shear capacity, respectively. The models provided by ACI 2013 \citep{american2013aci}, CNR-DT 215 2018 \citep{cnr215}, Triantafillou and Antonopoulos \citep{triantafillou2000design}, Triantafillou \citep{triantafillou1998strengthening} and Triantafillou \citep{triantafillou2016} are conservatively predicting lower than experimental values for the shear capacity of a TRM-strengthened URM wall, for the majority of the examples. On the other hand, the equations of CNR-DT 20 R1/2013 \citep{cnr200}, EN 1996 \citep{en1996}, EN 1998 \citep{en1998}, and Thomoglou et al. \citep{thomoglou2020ultimate} provide a more unbiased assessment of shear capacity with balanced positive and negative residuals. Among the existing models, this of Thomoglou et al. \citep{thomoglou2020ultimate} is the most effective, with a MAPE equal to 22.95\%. The proposed ML model outperforms all the others, presenting much lower prediction error based on the train and test MAPE (4.33\% and 8.03\%) and much less dispersion of the predictions around the identity line, as depicted in Fig.~\ref{comp_previous}i.

\subsection{Interpretation of the Proposed Model}
\begin{figure*}
\includegraphics[width=\textwidth]{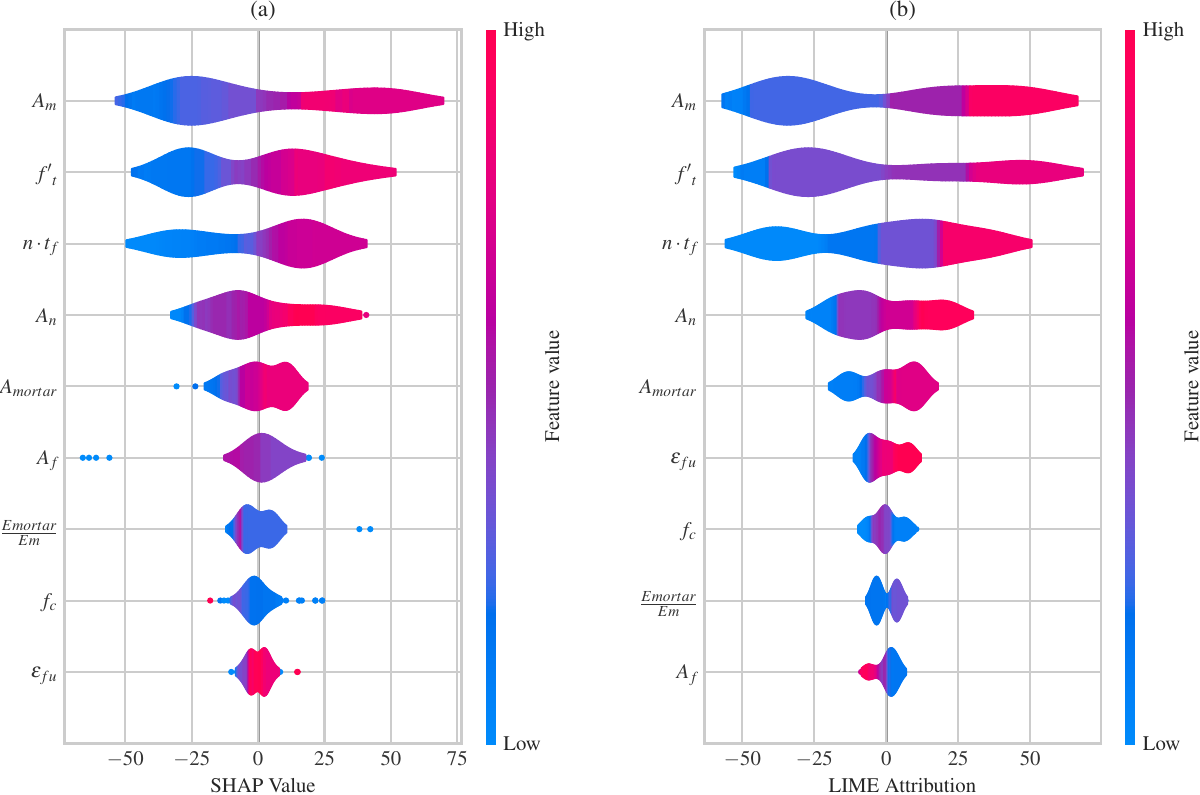}
\caption{(a) SHAP and (b) LIME violin plots}\label{shap_lime}
\end{figure*}

Explaining the prediction process is critical for models with applications in public safety, such as structural engineering problems. State of the art ML methods, such as ensemble methods and neural networks, have been criticized as "black boxes" that provide a limited understanding of their internal function. The utilization of ML interpretation techniques could enhance the reliability of the ML methods by providing insights into how a complicated model makes its predictions. Also, it is essential for the decision makers to have the potential to understand a model's predictions in order to trust its output values. Recently, techniques to address this issue have been introduced. These techniques are called interpretation or explanation techniques and are divided into local and global techniques. Local techniques aim to explain an individual prediction of an ML model and provide information about each input feature's contribution to the dependent variable's predicted value. On the other hand, global methods target the influence of each predictor on the general model trend. Also, another distinguishment of interpretation methods is between model agnostic and model-specific. The first could be applied to any ML method, while the last leverages the analysed model structure or parameters. In this section, six interpretation techniques are explored: SHapley Additive exPlanations (SHAP) \citep{shap}, Local Interpretable Model-Agnostic Explanations (LIME) \citep{lime}, Permutation Importance \citep{molnar2019interpretable}, Impurity-based Importance \citep{molnar2019interpretable}, Individual Conditional Expectation (ICE) \cite{ice} and Partial Dependence Plot (PDP) \citep{ale} which offer unique perspectives on model behaviour. 
Firstly, SHAP, a powerful, model-agnostic tool providing both global and local interpretability, is examined. SHAP values are based on the concept of Shapley values from cooperative game theory. They represent the average marginal contribution of a feature across all possible combinations of features. For a prediction model f and a specific instance x, the SHAP value for feature i is calculated as:
\begin{equation}\label{shap_eq}
\phi_i(x) = \sum_{S \subseteq T} \frac{|S|!(|T|-|S|)!}{|T|!}[f(x_S \cup {x_i}) - f(x_{S})]
\end{equation}
where $T$ is the set of all feature indices and $x_S$ is the input vector with only the features in $S$ present. The formula represents the average difference that adding feature $i$ to the input makes over all possible combinations of the remaining features. The distribution of the calculated SHAP values across the analysed dataset is shown in Fig.~\ref{shap_lime}a, in which it could be observed that the wider range of SHAP values exhibited by the h parameter parameters seems to have mainly positive influence in the prediction of shear capacity.

Next, LIME is a local and model-agnostic explanation method that, given a specific prediction, approximates the complex model locally with an interpretable one, like a linear regression or a single decision tree. More concretely, LIME uses the "black box" model to make an artificial dataset from its predictions in the vicinity of the examined point and, in sequence, trains the surrogate model on this dataset. As a result, LIME use the structure of the simpler model to extract the contribution of each predictor.

\begin{figure*}
\includegraphics[width=\textwidth]{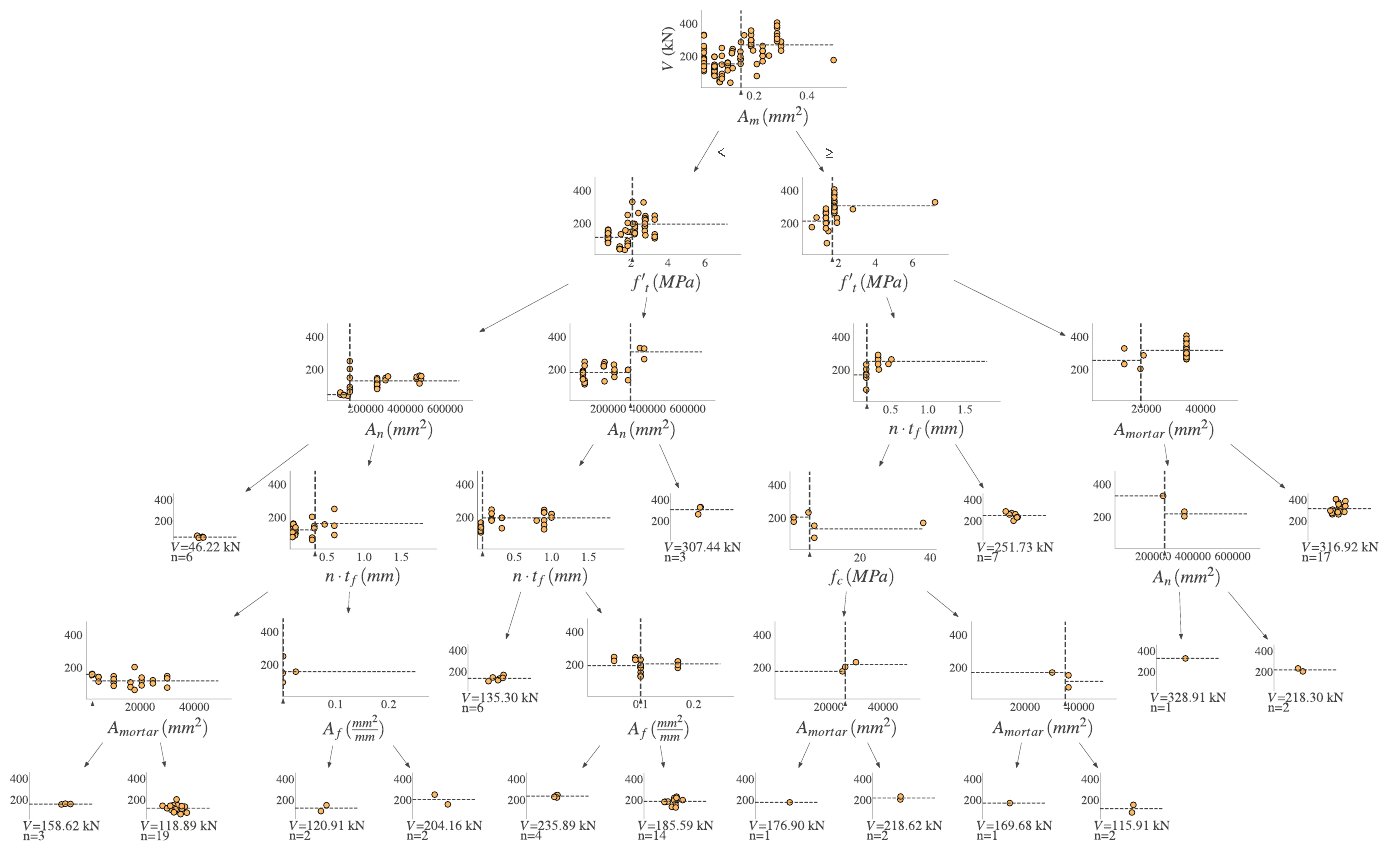}
\caption{Visualization of the first decision tree of the best XGBoost model}\label{dtree}
\end{figure*}

\begin{figure*}
\includegraphics[width=\textwidth]{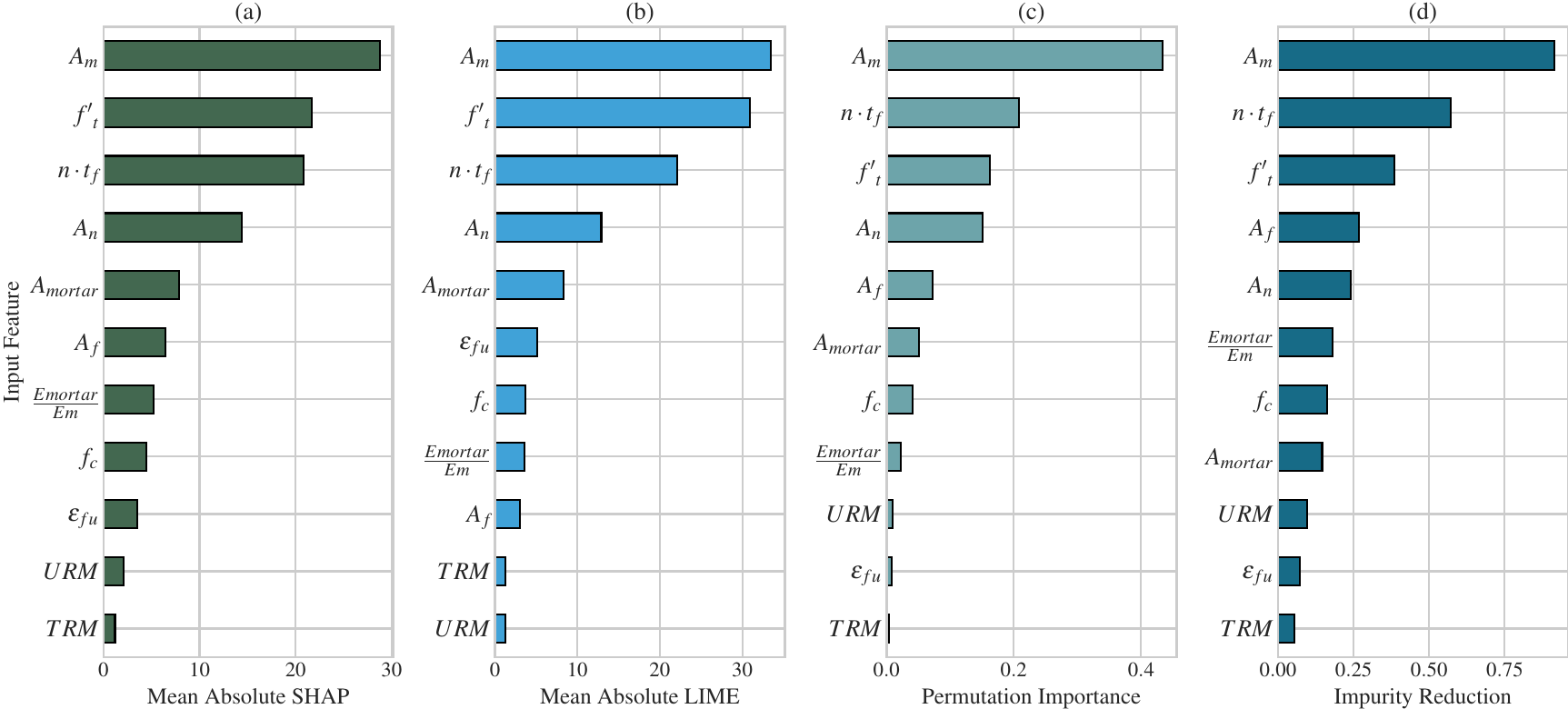}
\caption{Feature Importances according to (a) SHAP (b) LIME (c) Permutation and (d) Impurity methods}\label{feature_imps}
\end{figure*}

Moving to a more global perspective, the Permutation Importance evaluates each input feature's significance by randomly shuffling its values across the dataset examples and measuring the prediction performance's degradation. This method is a global interpretation and model-agnostic and serves as a straightforward and effective way to identify features that the model relies on most across the dataset. The Impurity Importance method (Equation~\ref{impurity_eq}) is model-specific and applicable to tree-based models like the XGBoost and CatBoost models.
\begin{equation}
Importance_i = Error_{permuted,i} - Error_{original}\label{impurity_eq}
\end{equation}
Fig.~\ref{feature_imps} presents the calculated feature importances according to the above described explanation and interpretation methods. In the first two subfigures (Fig.~\ref{feature_imps}(a,b)), the significance of the examined predictors is approximated through the mean absolute values of SHAP and LIME values, respectively. Both the two methods agreed for the five most meaningful input features are in descending order of significance: $A_m$, $f'_t$, $n\cdot t_f$, $A_n$, $A_{mortar}$. The Permutation and Impurity reduction methods are not essentially differentiated, indicating that the three most substantial predictors are $A_m$, $n\cdot t_f$, $f'_t$ and $A_n$, All the methods agreed that the most essential parameter is the $A_m$ followed by the pair of $f'_t$, $n\cdot t_f$. The above outcomes could also be verified from the visualization of the structure of the XGBoost model first Decision Tree in Fig~\ref{dtree}. The importance of each parameter relies on the proximity to the top of the tree and on the frequency with which this parameter is used as a splitting feature. As expected, the $A_m$ is placed at the top node as the most significant parameter, followed by $f'_t$ at the second level nodes of the tree. In the third level, the $n\cdot t_f$, $A_n$ and $A_{mortar}$ are employed as splitting features. The two most frequently occurring predictors in the decision tree are $n\cdot t_f$ and $A_n$, which confirms their participation at the most significant input variables.

\begin{figure*}
\includegraphics[width=\textwidth]{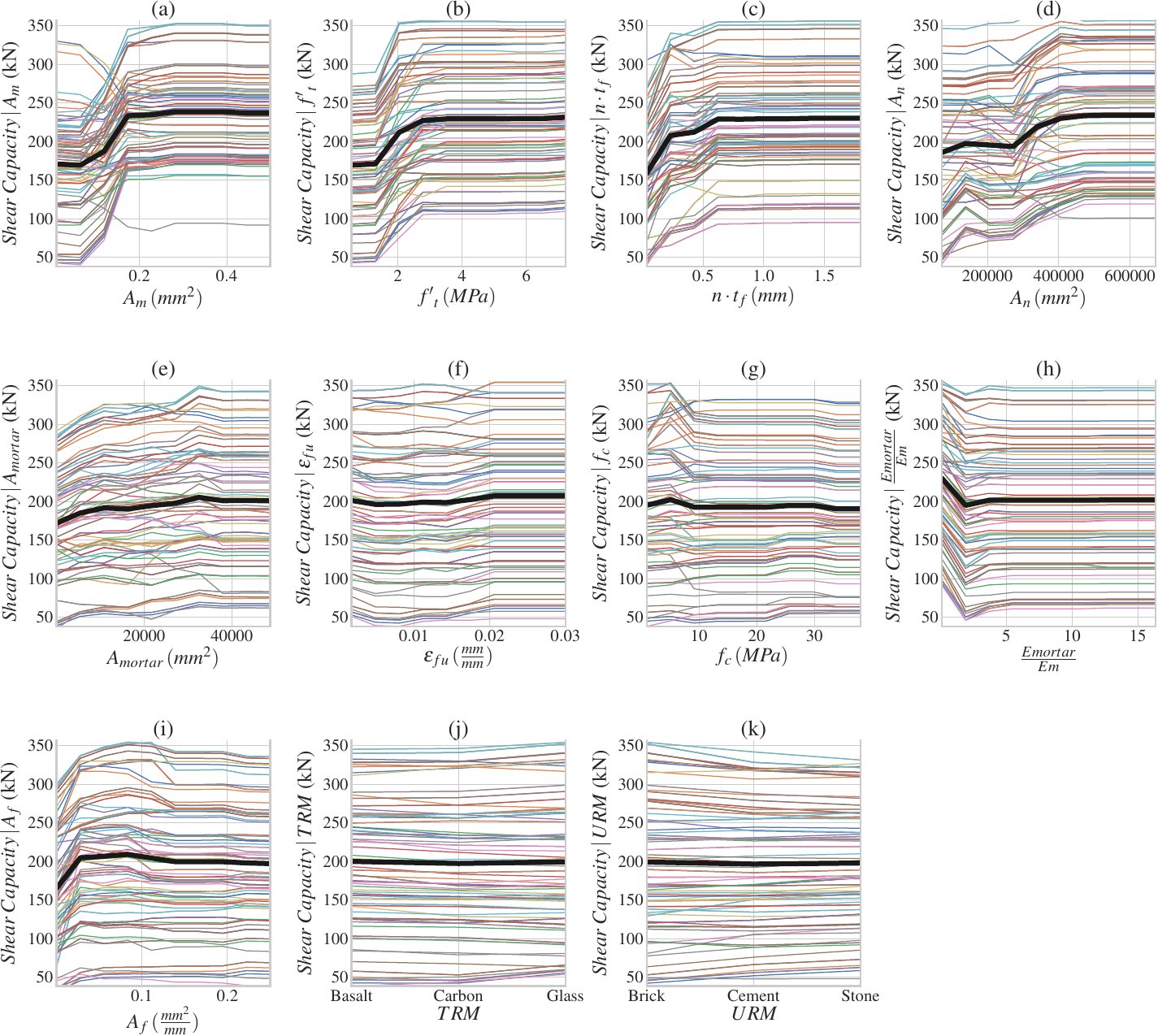}
\caption{The Individual Conditional Expectations (ICEs) plotted for each data point of the dataset with different colour and Partial Dependence Plot depicted using the black line for each input feature}\label{ice_pdp}
\end{figure*}

The ICE gives a separate function for a single data point, which describes the evolution of the prediction as the value of a single feature is modified. This represents the partial influence of an individual predictor on the prediction of a certain example. In practice, the ICE plot is created by evaluating the model multiple times, varying the feature of interest on a grid while holding all other features constant. Fig.~\ref{ice_pdp} shows the ICE plots with different colours for each example of our dataset. Each subplot illustrates the impact of only one predictor each time, including both the numerical and categorical ones. Continuing the PDP to operate from a more global perspective gives insights into the marginal impact of each input feature over the total dataset. In the same figure, the PDP is depicted with thick black lines, which average the ICEs at each value of the predictor's grid. The bigger the range of shear capacity values on the y-axis affects a predictor, the more important it is for the investigated problem. The most crucial input parameter seems to be the $A_m$ followed by the $f'_t$, the $n\cdot t_f$ and the $A_n$. $A_m$ has a zero average impact at the first part of PDP, an augmenting branch continues until approximately 0.2 $mm^2$, and finally, the partial derivative stays almost zero with small variations. Similar patterns are presented by the following features $f'_t$, the $n\cdot t_f$ and the $A_n$ in Fig.~\ref{ice_pdp}(b-d), which have mainly increasing effects in shear capacity with several horizontal and little decreasing branches.

\subsection{The Final Deployment and the Web-Application}
\begin{figure*}
\includegraphics[width=\textwidth]{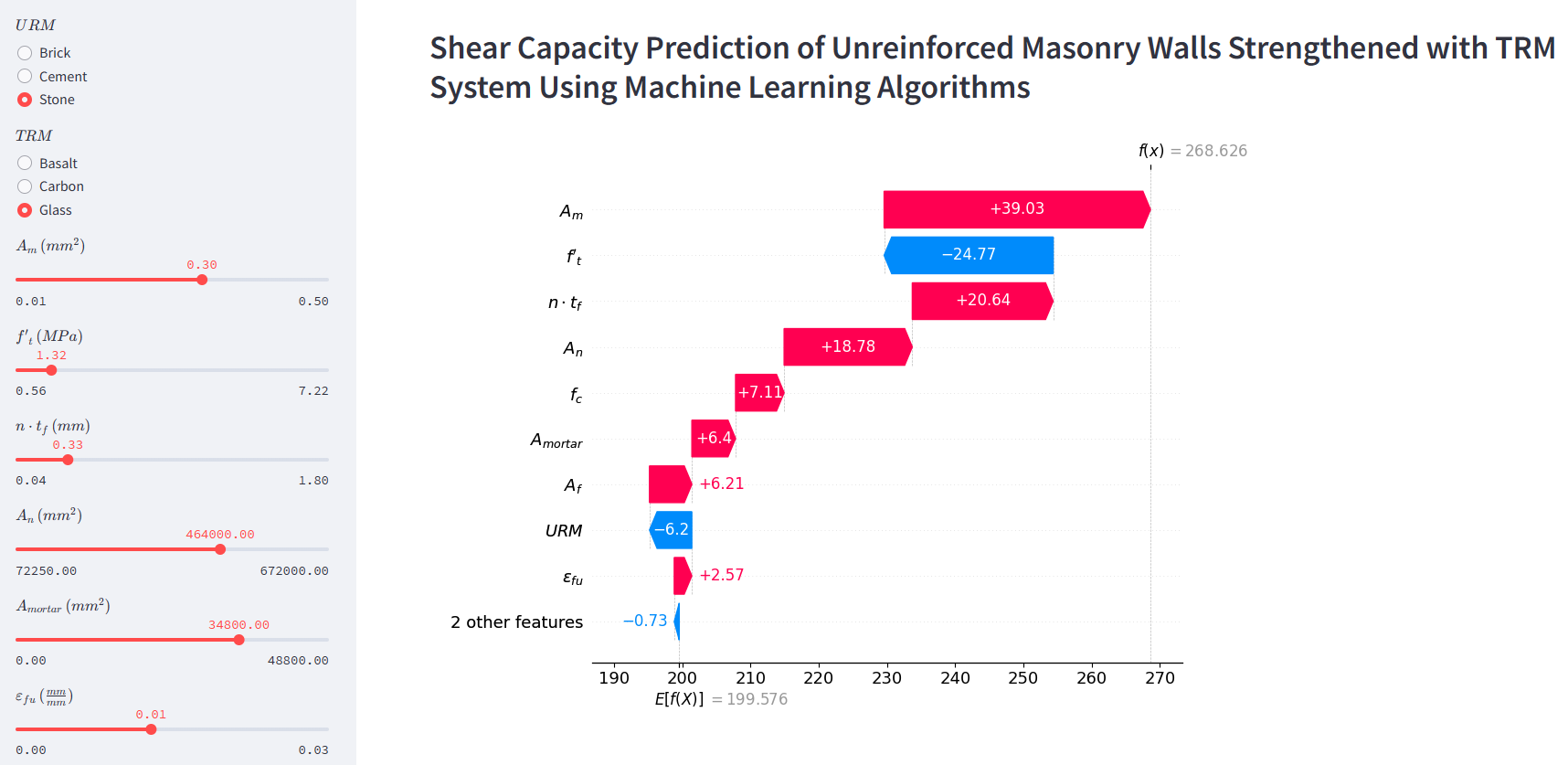}
\caption{The final deployment of the Machine Learning Methodology in a user-friendly Web-Application \href{https://trmwebapp.streamlit.app/}{https://trmwebapp.streamlit.app/}}\label{app}
\end{figure*}

Even though there is an extended application of ML in structural and earthquake engineering, only a few studies \citep{lazaridis2022applsci,wakjira2022explainable,abushanab2023machine,thisovithan2023novel,wakjira2024explainable} have implemented their methodologies in a ready-to-use tool. To make the results of the presented study accessible and valuable to both researchers and professionals, the configured machine learning system is deployed in a user-friendly web application. This application delivered the outcome of our study in an easy-to-access graphical interface and took advantage of the local interpretation method SHAP to explain the contribution of each input parameter to the prediction of shear capacity. The web application was implemented using the Streamlit \citep{khorasani2022streamlit} Python library, and a snapshot of it is depicted in Fig.\ref{app}. The regressor was trained on the entire dataset to utilize as much data as possible, as is common in real-world machine learning systems. 

\section{Conclusions}
This study is a novel investigation of an experimental dataset selected after a bibliographic search, using machine learning regression and interpretation methods to build a reliable and practical tool for the shear capacity prediction of a URM wall strengthened with TRM jackets. The raw data was processed using one-hot encoding for the categorical features and the standard scaling for the numerical variables; in continuing, the outliers of the dataset were recognized and removed. The processed dataset was divided into train and test datasets. The hyperparameter tuning was conducted in the training dataset using the k-fold strategy with k equal to 5. The optimized model of each method was evaluated in the test set, and the performances of all models were compared using five different metrics. The two most effective methods were found to be the XGBoost and CatBoost methods. A blended model which averages the predictions of the above exhibited the best prediction capacity according to all the evaluation metrics. The best model was interpreted using local and global interpretation techniques and deployed in a web application. The main conclusion of the present study could be summarised as follows:
\begin{itemize}
    \item The best prediction ability on the test set exhibited by XGBoost and CatBoost tuned models with $R^2$ equal to 0.94 and 0.93 and MAPE equal to 9.3\% and 10.5\%, respectively. Even better performance is obtained if the predictions of the XGBoost and CatBoost models are combined using a Voting Regressor. The consequent Voting model emerged as the most efficient of all, achieving an $R^2$ equal to 0.95 and MAPE equal to 8.03\% on the test set.
    \item The importance analysis using SHAP, LIME, Permutation, and Impurity Reduction interpretation methods revealed that $A_m$, $f_t$, and $n\cdot t_f$ are the most important predictors for estimating the TRM-strengthened shear capacity of masonry walls. Furthermore, the PDP as a global interpretation technique provides useful insights into the marginal effect of each predictor on the shear capacity.
    \item As the more practical outcome of this study, the proposed methodology was implemented in a user-friendly web application (\href{https://trmwebapp.streamlit.app/}{https://trmwebapp.streamlit.app/}) which both researchers and professionals could employ as a reliable, speedy and easy to use tool for the shear capacity estimation of TRM-strengthened masonry walls. 
\end{itemize}

\bibliographystyle{unsrt} 
\bibliography{references.bib}
\end{document}